\documentclass[twocolumn,showpacs,amsmath,amssymb,prd]{revtex4}
\usepackage{amsmath}\usepackage{amssymb}\usepackage{amsfonts}
\usepackage{natbib}\usepackage{graphics}\usepackage{latexsym}\usepackage{dcolumn}\usepackage{bm}
\usepackage[dvips]{epsfig}
\usepackage{color}

\begin{document}
\setlength{\unitlength}{1mm}
\newcommand{\CB}{{\cal B}}
\newcommand{\CD}{{\cal D}}
\newcommand{\CQ}{{Q}}
\newcommand{\CE}{{\cal E}}
\newcommand{\CW}{{\cal W}}
\newcommand{\CR}{{R}}
\newcommand{\CH}{{\cal H}}
\newcommand{\CM}{{\cal M}}
\newcommand{\CS}{{S}}
\newcommand{\CP}{{P}}
\newcommand{\CC}{{\cal C}}
\newcommand{\CF}{{F}}
\newcommand{\Omalpha}{{\Omega_{\alpha}}}
\newcommand{\Ombeta}{{\Omega_{\beta}}}
\newcommand{\Omgamma}{{\Omega_{\gamma}}}
\newcommand{\average}[1]{\left\langle #1 \right\rangle_\CD}
\newcommand{\initial}[1]{{#1_{\rm \bf i}}}
\title{Effective Inhomogeneous Cosmologies and Emerging Scalar Fields\footnote{Contribution presented at the workshop {\it New Directions in Modern Cosmology},
Leiden, The Netherlands, 27.9. -- 1.10., 2010.}}
\author{Thomas Buchert$^{1}$, Nathaniel Obadia$^{2}$ and Xavier Roy$^{1}$}
\affiliation{$^{1}$Universit\'e Lyon~1, Centre de Recherche Astrophysique de Lyon,
\\9 Avenue Charles Andr\'e, F--69230 Saint--Genis--Laval,
France}
\affiliation{$^{2}$\'Ecole~Normale~Sup\'erieure de Lyon, Centre de Recherche
Astrophysique de Lyon,\\
46 All\'ee d'Italie, F--69364 Lyon Cedex 07, France}
%
%


\begin{abstract}
In this contribution we summarize two recent applications of a correspondence between backreaction terms in averaged inhomogeneous cosmologies and an effective scalar field (the ``morphon'').
Backreaction terms that add to the standard sources of Friedmannian kinematical laws and that emerge from geometrical curvature invariants built from inhomogeneities, can be interpreted in terms of a minimally coupled scalar field in the case of a dust matter source. We consider closure conditions of the averaged equations that lead to different evolution scenarii: a) the standard Chaplygin equation of state imposed as an effective relation between kinematical fluctuations and intrinsic curvature of space sections, and b) an inflationary scenario that emerges out of inhomogeneities of the Einstein vacuum, where averaged curvature inhomogeneities model the potential of an effective classical inflaton.  
\end{abstract}

\maketitle

\section{Introduction} 

Our universe is supposed to verify the strong cosmological principle which demands homogeneity and isotropy at all scales. This standard approach, known as Friedmann--Lema\^{\i}tre--Robertson--Walker (FLRW) cosmology obeys the set of equations,
\begin{subequations}
\begin{eqnarray}
	\left(\frac{\dot a}{a}\right)^2 = \, \frac{8 \pi G}{3} \sum_i \varrho_{(i)} \, - \, \frac{k}{a^2} \, , \\
	\frac{\ddot a}{a} \, = \, - \frac{4 \pi G}{3} \sum_i \left(\varrho_{(i)} + 3 p_{(i)}\right) \, , \\
	\dot{\varrho}_{(i)} \, + \, 3 \, \frac{\dot a}{a} \left( \varrho_{(i)} +  p_{(i)}\right) \, = \, 0 \, , \label{eq:hom_dyn}
\end{eqnarray}
\end{subequations}
that, together with equations of state as relations between the homogeneous variables, e.g. between the pressures and energy densities, forms a closed system.
The FLRW framework is widely used in order to describe the dynamics of our Universe and the formation of its constituents. It, however, leaves in suspense an explanation about the origin of Dark Energy and Dark Matter, which respectively represent in this model about 3/4 and 1/4 of the total content of the universe model. This last point might actually reveal a symptom of a deeper problem linked to this approach. Indeed, in FLRW cosmology one determines background quantities regardless of the scale and makes them evolve according to a homogeneous--isotropic solution of Einstein's equations. But, are the background quantities well--defined within standard cosmology, {\it i.e.} as a suitable average over the inhomogeneities? Is their evolution well--approximated in this framework, {\it i.e.} is the time dependence of the homogeneous--isotropic averaged state well approximated by a homogeneous--isotropic solution? All these issues can be routed back to the nonlinearity of the gravitational equations and the averaging problem \cite{ellis:average,ellisbuchert,buchert:review}.

Rewriting Einstein's equations via the ADM formalism and spatially averaging, in a domain--dependent way, the scalar parts of the equations with respect to free--falling fluid elements, the averaged dynamics of an inhomogeneous universe model filled with a pressureless fluid assumes the following form \cite{buchert:dust,buchert:fluid,buchert:review}:
\begin{subequations}
\begin{eqnarray}
	\left(\frac{{\dot a}_\CD}{a_\CD}\right)^2 = \, \frac{8 \pi G}{3} \average{\varrho} \, - \, \frac{k_\initial\CD}{a_\CD^2} - \frac{\CW_\CD + {\CQ}_\CD}{6} \, , \\
	\frac{{\ddot a}_\CD}{a_\CD} \, = \, - \frac{4 \pi G}{3} \average{\varrho} \, + \, \frac{{\CQ}_\CD}{3} \, , \\
	\langle{\varrho}\rangle\dot{}_\CD \,+ \, 3 \, \frac{{\dot a}_\CD}{a_\CD} \average{\varrho} \, = \, 0 \, , \\
	\frac{1}{a_\CD^6} \,  \left(\,{\CQ}_\CD \, a_\CD^6 \,\right)\dot{} \, + \, \frac{1}{a_\CD^{2}} \, \left(\CW_\CD \, a_\CD^2 \, \right)\dot{} \, = \, 0 \, , \label{eq:inhom_dyn}
\end{eqnarray}
\end{subequations}
where $a_\CD$ is the effective scale factor, $\average{\varrho}$ is the energy density of the irrotational dust averaged over a compact, mass--preserving domain $\CD$, $\CQ_\CD$ is the kinematical backreaction term (an extrinsic curvature invariant), and finally $\CW_\CD$ is the curvature deviation from a constant--curvature $k_{\initial\CD}$ according to the FLRW solution (an intrinsic curvature invariant). These variables are defined as
\begin{subequations}
\begin{eqnarray}
	\average{\varrho}(t) = \frac{M_{\initial\CD}}{V_{\initial\CD}}  \, a_\CD^{-3}  \,\, , \,\, a_\CD (t) := \left( \frac{V_{\CD}(t)}{V_{\initial\CD}} \right)^{1/3} \,, \\
	{\CQ}_\CD (t) := \frac{2}{3}\average{\left(\theta - \average{\theta}\right)^2 } - 2\average{\sigma^2} \, , \\
	\CW_\CD (t)  := \average{\CR} - \frac{6 k_\initial\CD}{a_\CD^2}\, ,
\end{eqnarray}
\end{subequations}
with $V_{\initial\CD}$ the initial volume of the domain and $V_{\CD}(t)$ its volume at a proper time $t$, $\theta$ the rate of expansion, $\sigma$ the rate of shear, and $R$ the $3-$Ricci scalar curvature. Comparing the set of equations (1) and (2), one easily notices that the average evolution of an inhomogeneous universe model differs from the evolution of a homogeneous one. The change of the cosmological background evolution is here driven by the non--trivial geometrical structure of an inhomogeneous space, and the corresponding deviations are encoded into the backreaction terms $\CQ_\CD$ and $\CW_\CD$, which are coupled through the relation (\ref{eq:inhom_dyn}).

Other formulations of the effective inhomogeneous equations can be done. The first one aims at considering the backreaction terms as an effective fluid by introducing 
an effective perfect fluid energy momentum tensor with
\begin{equation}
\varrho^{\CD}_{b} = - \frac{1}{16\pi G} ({\CQ}_\CD + \CW_\CD )\;;\; p^{\CD}_{b} = - \frac{1}{16\pi G} ({\CQ}_\CD - \frac{1}{3} \CW_\CD ), 
\label{backreactionfluid}
\end{equation}
leading to the following reformulation of the system (2):
\begin{subequations}
\begin{eqnarray}
	\left(\frac{{\dot a}_\CD}{a_\CD}\right)^2 \, = \, \frac{8 \pi G}{3} \, \left( \average{\varrho} + \varrho^{\CD}_{b} \right) \, - \, \frac{k_{\CD_{\it i}}}{a^2_\CD} \, , \\
	\frac{{\ddot a}_\CD}{a_\CD} = \, - \frac{4 \pi G}{3} \, \left( \average{\varrho} + \varrho^{\CD}_{b} + 3 {p}^{\CD}_{b} \right) \, , \\
	\langle{\varrho}\rangle\dot{}_\CD \,+ \, 3 \, \frac{{\dot a}_\CD}{a_\CD} \average{\varrho} \, = \, 0 \, , \\
	{\dot\varrho}^{\CD}_{b} + 3 \, \frac{{\dot a}_\CD}{a_\CD} \, \left(\varrho^{\CD}_{b} + {p}^{\CD}_{b} \right) = 0 \, ,
\end{eqnarray}
\end{subequations}
where we see that the coupling between the backreaction terms now stands for the conservation law of the backreaction fluid. We also appreciate that, like in the standard model,
we need an equation of state to close the system that here is dynamical and furnishes a relation between the effective sources. We shall consider two of such equations of state employed as closure conditions for the set of equations (5).

The second reformulation, suggested by the form of the effective sources (\ref{backreactionfluid}) \cite{buchert:static}, consists in describing the backreaction fluid by a minimally coupled real scalar field $\phi_\CD$,~called the {\it morphon field}, evolving in an effective potential $U_\CD(\phi_\CD)$ \cite{morphon}: 
\begin{equation}
\varrho_{b}^{\CD} = \varrho_{\phi}^{\CD} : = \epsilon \dot{\phi}^2_{\CD}/2 + U_{\CD}\;;\;p_{b}^{\CD} = p_{\phi}^{\CD} : = \epsilon \dot{\phi}^2_{\CD}/2 - U_{\CD},
\end{equation}
where $\epsilon = + 1$ for a standard scalar field (with a positive kinetic energy), and $\epsilon = - 1$ for a phantom scalar field (with a negative kinetic energy). The scalar field language leads to the following {\it correspondence relations}:
\begin{subequations}
\begin{eqnarray} 
	\epsilon \dot{\phi}^2_{\CD} = - \frac{1}{8 \pi G} \, ({\CQ}_\CD + \frac{1}{3}{\CW}_\CD ) \, , \\
	U_{\CD} = - \frac{1}{24 \pi G}  {\CW}_\CD \, , \\
	\ddot{\phi}_{\CD} + 3 \frac{\dot{a}_\CD}{a_\CD} \, \dot{\phi}_{\CD} + \epsilon \frac{\partial}{\partial \phi_{\CD}} U_{\CD} = 0 \, ,
	\label{eq:morphon}
	\end{eqnarray}
\end{subequations}
where the Klein--Gordon equation is simply the counterpart of the conservation law for the backreaction fluid. This correspondence allows us to interpret the kinematical backreaction effects in terms of the properties of scalar field cosmologies, notably quintessence or phantom--quintessence scenarii that are here routed back to models of inhomogeneities.
\section{Backreaction fluid as a Chaplygin gas}

In the above introduced effective inhomogeneous cosmologies, the dark components may be unified through the kinematical backreaction: $\CQ_\CD < 0$ effectively mimics Dark Matter and $\CQ_\CD > 0$ effectively mimics Dark Energy. Both Dark Energy and Dark Matter are then not included as additional sources but are both manifestations of spatial geometrical properties.

Another unification of the dark components is realized thanks to the standard Chaplygin gas (CG). This exotic fluid, whose state equation reads $p = - A / \varrho$, with $A > 0$, has the interesting property to connect Dark Matter and Dark Energy behaviors via its evolution \cite{gor:demodel1,gor:demodel2}.

The two previous features motivate us to build a model in which the backreaction fluid is supposed to obey a scale--dependent CG equation of state: $p_b^\CD = - A_\CD / \varrho_b^\CD$ with $A_\CD > 0$ \cite{chaplygin}. This construction allows to unify the dark components, first, simultaneously, thanks to the scale dependence of the backreaction fluid, and, second, through its evolution. 

The origin and the magnitude of the Dark Energy (first situation in Figure 1) and of both Dark Energy and Dark Matter (second situation in Figure 1) can be entirely explained by the particular geometrical structure of an inhomogeneous space obeying the Chaplygin equation of state.

In our framework, it is necessary to reinterpret observational data. Indeed, angular diameter and luminosity distances, for instance, depend on metrical properties related to the averaged scalar curvature which evolves differently compared with its Friedmannian counterpart. In light of this remark it is premature to exclude \cite{wmapchap,generalchap} the standard Chaplygin equation of state as providing a good match with observational data.

Finally, note also that, depending on the initial conditions of the kinematical backreaction, the volume of the domain $\CD$ might undergo different dynamics: its expansion is first decelerated then accelerated; its expansion is always accelerated; or its expansion is accelerated, then decelerated and again accelerated \cite{chaplygin}. The latter case is an interesting situation since two accelerated phases occur. Our model only concerns the matter--dominated universe; however, this situation allows us to imagine that the accelerated phase in primordial inflation and the one occurring today might be driven by the same mechanism that has to be studied in a more general model. Indeed, we are going to explain the first steps towards a modeling of Inflation in the next section.

\begin{figure}[h!]
		\includegraphics[width=8.5cm]{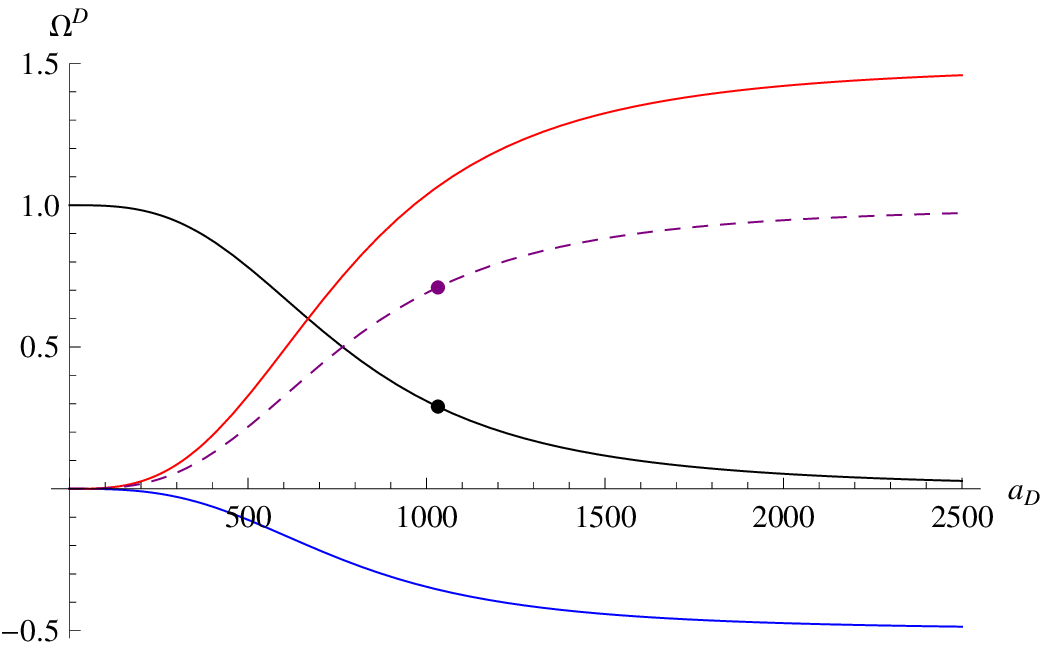}
		\includegraphics[width=8.5cm]{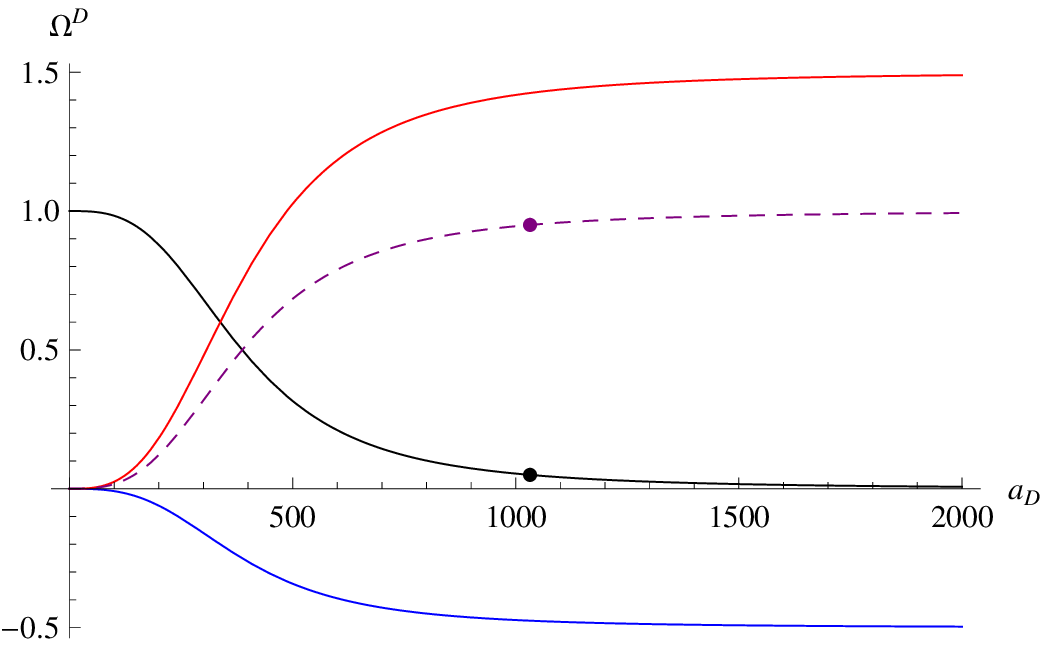}
\caption{\cite{chaplygin} Evolution of the domain--dependent cosmological parameters $\Omega_m^{\CD}$ (black), $\Omega_{\CW}^{\CD}$ (red), $\Omega_{\CQ}^{\CD}$ (blue) and $\Omega_X^{\CD} = \Omega_{\CW}^{\CD} + \Omega_{\CQ}^{\CD}$ (dashed) w.r.t. the effective scale factor.\\ Upper figure: $\Omega_X^{\CD}(a_\CD \sim 1000) = \Omega_{DE}^0 \sim 0.72$ (purple dot) and $\Omega_m^{\CD}(a_\CD \sim 1000) = \Omega_b^0 + \Omega_{DM}^0 \sim 0.28$ (black dot).\\ Lower figure: $\Omega_X^{\CD}(a_\CD \sim 1000) = \Omega_{DE}^0 +  \Omega_{DM}^0 \sim 0.95$ (purple dot) and $\Omega_m^{\CD}(a_\CD \sim 1000) = \Omega_b^0 \sim 0.05$ (black dot).}
\end{figure}
\section{Inflation from Inhomogeneities}
The effect of inhomogeneities can play a role in {\it primordial} cosmology too, when using the morphon description. Indeed, Inflation, the most prevalent paradigm used to cure the flatness, smoothness and horizon issues of the standard model of cosmology, is often described through 
the dominance of a slow--rolling scalar field, namely the inflaton, 
over the other components of the energy budget. 
Furthermore, there is a general trend relying on the FLRW description to assume Inflation unsustainable when the initial conditions are even slightly inhomogeneous, contrary to the idea that Inflation should stem from {\it chaotic} initial conditions.

We offer the idea \cite{inflation1} that the morphon can formally play the role of an effective inflaton 
and that inhomogeneities could be the actual cause of a de Sitter--like era prior to the last scattering surface. 
In this case, several interpretations emerge and differ from ordinary Chaotic Inflation:

\begin{itemize}
 \item the nature of the inflaton is solved;
 \item the inflaton/morphon is purely classical, therefore  its value is no longer bounded by $M_{Pl}$,
 and Eternal Inflation is naturally avoided;
 \item the inflaton's wave equation does not come from an additional term in the action but is rather given by the morphon's integrability condition (\ref{eq:inhom_dyn}., \ref{eq:morphon});
 \item the choice of the potential is forced by physical requirements such as
\vspace{-5pt}
  \begin{itemize}
    \item a true initial conditions' freedom of choice ($\CQ_\initial\CD \lessgtr 0$, see Figure 2),
    \item the necessity to end the expansion at one point,
    \item the necessity to describe the vanishing of inhomogeneities at that end,
  \end{itemize}
 \item the emergence of Inflation within a (physical) inhomogeneous background is allowed/favored.
\end{itemize}
\noindent
Motivated by these requirements we choose the Ginzburg--Landau potential,
\begin{equation}
 U_\CD^{GL} = \frac{\lambda}{4} \left(\phi_\CD^2 - \frac{M_\phi^2}{\lambda} \right)^2
=\frac{M_\phi^4}{4\lambda} - \frac{1}{2} M_\phi^2 \phi_\CD^2 + \frac{\lambda}{4} \: \phi_\CD^4 \, ,
\end{equation}
as a sufficiently generic and conservative example of an initial conditions poorly dependent potential,
while the kinetic part of the morphon is canonical ($\epsilon=+1$).
 
\begin{figure}[h!]
\includegraphics[width=7cm]{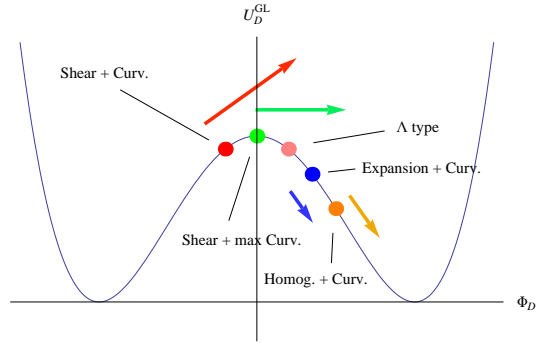}
\caption{\cite{inflation1} Reinterpretation of the initial conditions in the language of 
the backreaction quantities: the Ginzburg--Landau potential in arbitrary units and the possible 
initial conditions as well as their physical meaning.
All conditions possess some curvature $\CW_\CD^i<0$.
The arrows schematically indicate the amplitude of the morphon's initial speed $\dot\Phi_\CD^i$. 
In the order of the points (from left to right): the first two points dominated by shear fluctuations (red, green) are obtained for 
$\CQ_\CD^i < 0 \Leftrightarrow \dot\Phi_\CD^{i\:2} > 2(H_\CD^{i\:2}+ k_\CD^i )$; 
the next points dominated by expansion fluctuations (blue, pink) for $ \dot\Phi_\CD^{i\:2} < 2(H_\CD^{i\:2}+ k_\CD^i )$,
where the de Sitter--$\Lambda$ equivalent case has a stiff morphon $\dot\Phi_\CD^i = 0$; 
the homogeneous case (last point, orange) is obtained for  $ \dot\Phi_\CD^{i\:2} = 2(H_\CD^{i\:2}+ k_\CD^i )$.}
\end{figure}

The two parameters $\lambda$ and $M_\phi$ both depend on the average size of the Hubble radius at which Inflation takes place and fix the number of e--folds the latter lasts.

\noindent
We consider the energy budget in the averaged models,
\begin{subequations}
\begin{eqnarray}
\Omega_m^\CD + \Omega_k^\CD+ \Omega_\CW^\CD + \Omega_\CQ^\CD = 1\;, \,\\
\Omega_m^\CD=\frac{8\pi G \average{\rho}}{3H_\CD^2}\;,\quad
\Omega_k^\CD=-\frac{k_\CD}{a_\CD^2H_\CD^2}\;,\nonumber \\
\Omega_\CW^\CD=-\frac{\CW_\CD}{6H_\CD^2}\;,\quad
\Omega_\CQ^\CD=-\frac{\CQ_\CD}{6H_\CD^2}\;, 
\end{eqnarray}
\end{subequations}
and show in Figure 3 how, even in the absence of matter ($\Omega_m^\CD=0$), 
the energy density in the Friedmannian curvature $\Omega_k^\CD$ strongly decays and converts into the backreaction terms, 
while 
\begin{equation}
\epsilon_\CD  := - \dot H_\CD/H_\CD^2 = 1 + 
2 \Omega_\CQ^\CD \;<\; 1
\end{equation}
is the condition tantamount to Inflation.
In the presence of pressure--less matter, the same scenario occurs, irrespective of the initial conditions.

Related publications will soon appear on the the various attempts we made 
to render Inflation implicit and duration natural in inhomogeneous cosmologies.
We shall consider the interaction of the morphon with radiation, dust, and a fundamental scalar field, and we shall 
address the horizon problem. Another analysis is dedicated to the global instability of the FLRW backgrounds employing a
dynamical system analysis of the averaged equations in the phase space of the cosmological parameters.
The application to scalar field models for Dark Matter forms the subject of a further parallel attempt.

\begin{figure}
\includegraphics[width=8.5cm]{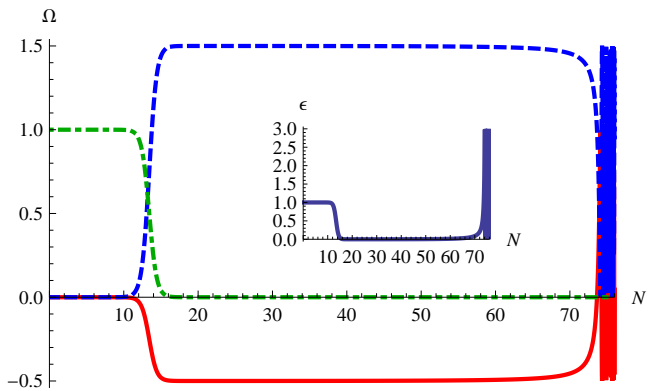}
\caption{\cite{inflation1} The energy densities $\Omega_\CD^\CQ$ (solid, red), 
$\Omega_\CD^\CW$ (dashed, blue), 
and $\Omega_\CD^k$ (dotdashed, green)  
for one of the cases of Fig.~2 as a function of the number of e--folds $N=\ln a_\CD$.
Here we considered the effect of inhomogeneities of curvature only, {\it i.e.} the matter density is assumed to be zero.
The initial value of the homogeneous part of the intrinsic curvature  
has been taken arbitrary large (e.g. $\Omega_\CD^{k_i} \simeq 1$) 
to underline the fact that it vanishes anyway after a few e--folds. 
In the inset, the slow--roll parameter $\epsilon_\CD$ for the same configuration is shown.
Since $\epsilon_\CD =\Omega_\CD^k + \Phi_\CD'^2/2$ always holds, 
$\epsilon\simeq 1$ in the pre--inflation era when $\Omega_\CD^k\gg\Omega_\CD^\CQ,\Omega_\CD^\CW$,
and $\epsilon \simeq 0$ during Inflation when $U_\CD^{GL}\gg \Phi_\CD'^2/2$.\label{fig:epsilon}
\label{fig:omega}}
\end{figure}

\bigskip
\noindent
{\em Acknowledgments.} \\ We would like to thank Francesco Sylos--Labini and Giovanni Marozzi for interesting discussions during the workshop.
This work is supported by ``F\'ed\'eration de Physique Andr\'e--Marie Amp\`ere''. 


\begin{thebibliography}{2010}

\bibitem{buchert:review}
T. Buchert: Dark Energy from structure -- a status report.
Gen. Rel. Grav. {\bf 40}, 467 (2008)
%
\bibitem{ellis:average}
G.F.R. Ellis: Relativistic cosmology: its nature, aims and problems.
In \textit{General Relativity and Gravitation} (D. Reidel Publishing Company, Dordrecht), pp. 215--288 (1984)
%
\bibitem{ellisbuchert} G.F.R. Ellis and T. Buchert: The Universe
seen at different scales. \emph{Phys. Lett. A.} (Einstein Special Issue)
\textbf{347}, 38 (2005)
%
\bibitem{buchert:dust}
T. Buchert: On average properties of inhomogeneous fluids in general relativity: 1. dust cosmologies.
Gen. Rel. Grav. {\bf 32}, 105 (2000)
%
\bibitem{buchert:fluid}
T. Buchert: On average properties of inhomogeneous fluids in general relativity: 2. perfect fluid cosmologies.
Gen. Rel. Grav. {\bf 33}, 1381 (2001)
%
\bibitem{buchert:static}
T. Buchert: On globally static and stationary cosmologies with or without a cosmological constant and the dark energy problem.
Class. Quant. Grav. {\bf 23}, 817 (2006)
%
\bibitem{morphon}
T. Buchert, J. Larena and J.--M. Alimi: Correspondence between kinematical backreaction and scalar field cosmologies -- the 'morphon field'.
Class. Quant. Grav. {\bf 23}, 6379 (2006)
%
\\
\bibitem{gor:demodel1}
V. Gorini, A. Kamenshchik and U. Moschella: Can the Chaplygin gas be a plausible model for Dark Energy?
Phys. Rev. D {\bf 67}, 063509 (2003)
%
\bibitem{gor:demodel2}
V. Gorini, A. Kamenshchik, U. Moschella and V. Pasquier: The Chaplygin gas as a model for Dark Energy.
In: Proc. MG10, Rio de Janeiro, Brazil, 20-26 July 2003, M. Novello, S. Perez Bergliaffa, R. Ruffini (eds.). World Scientific, p.~840 (2005)
%
\bibitem{chaplygin} X. Roy and T. Buchert: Chaplygin gas and effective description of inhomogeneous universe models in general relativity. 
\emph{Class. Quant. Grav.} \textbf{27}, 175013 (2010).
%
\bibitem{wmapchap}
L. Amendola, F. Finelli, C. Burigana and D. Carturan: WMAP and the generalized Chaplygin gas.
JCAP {\bf 0307}, 005 (2003)

\bibitem{generalchap} M.~C. Bento, O. Bertolami and A.~A. Sen: Generalized Chaplygin gas model: dark energy -- dark matter unification and CMBR constraints.
Gen. Rel. Grav. {\bf 35}, 2063 (2003)

\bibitem{inflation1} T. Buchert and N. Obadia: Effective Inhomogeneous Inflation: curvature inhomogeneities of the Einstein vacuum, submitted. arXiv:1010.4512 (2010)
%

\end{thebibliography}
\end{document}